\documentclass{aa}  

\usepackage{graphicx,url}
\usepackage[dvipsnames]{xcolor}
\usepackage[normalem]{ulem}
\usepackage{mathrsfs,amssymb,amsmath}
\usepackage[varg]{txfonts}
\usepackage{booktabs}
\usepackage{multirow}
\usepackage{lineno}
\usepackage{siunitx} 
\usepackage[breaklinks, colorlinks, citecolor=blue]{hyperref}
\begin{document}

 \title{Investigating signatures of extinct PeVatron activity in supernova remnants IC~443 and W51~C}

   \author{S. Celli \inst{1,2}
          \and
          G. Morlino \inst{3}
          }
   \institute{Sapienza Universit\`a di Roma, Physics Department, P.le Aldo Moro 5, 00185, Rome, Italy
             \and
             Istituto Nazionale di Fisica Nucleare, Sezione di Roma,
             P.le Aldo Moro 5, 00185, Rome, Italy\\
             \email{silvia.celli@roma1.infn.it}
             \and
             INAF Osservatorio Astrofisico di Arcetri, Largo Enrico Fermi, 5, 50125, Florence, Italy\\
             \email{giovanni.morlino@inaf.it}
}

   \date{Received September XX ; accepted March XX}

  \abstract
   {Supernova remnants (SNRs) are well-known sources of non-thermal particles and likely the main contributors of Galactic cosmic rays (CRs). Recent ultra-high energy (UHE) gamma-ray data from middle-aged SNRs, such as IC~443 and W51~C, highlight the key role of these systems for particle acceleration. Understanding the SNR contribution to PeV CRs requires detailed modelling of how accelerated particles escape their shocks and are released inside the Galaxy.}
   {In this work, our aim is to critically assess the SNR paradigm for the origin of Galactic CRs by investigating the expected secondary radiation produced by hadronic collisions of confined and escaping protons in SNRs, and we compared it to recent UHE observations. }
   {Through a time-dependent analytical model describing particle acceleration at the SNR forward shock, we investigated particle escape from the shock expanding into a uniform environment and diffusing inside and outside the SNR.}
   {Our results support the interpretation that the gamma-ray emission from W51~C and IC~443 is due to hadronic emission from both confined and escaping protons, allowing us to derive robust constraints on the particle maximum energy, the magnetic field amplification process, and the CR acceleration efficiency in these systems.}
   {We find that the acceleration efficiency is in the range of $1-10\%$ ($2-12\%$) when the spectrum of accelerated particles is $\propto p^{-4}$ ($\propto p^{-4.3}$) and that the local diffusion coefficient is suppressed by a factor of three to ten compared to the average Galactic value. On the other hand, the particle maximum energy achieved falls, by a factor of a few, below the PeV energy required to explain the CR knee in the context of SNRs only.}


   \maketitle
%

\section{Introduction}
\label{sec:intro}
Supernova remnants (SNRs) have long been considered the major sources of Galactic cosmic rays (CRs) because multi-wavelength observations suggest protons and electrons are energized in these systems \citep{fermiPB2013}. The so-called SNR paradigm for the origin of Galactic CRs relies on the fact that particle acceleration, as due to diffusive shock acceleration (DSA; \citealt{axford,kr,bell78,blandford87}), requires effective confinement of particles in the shock region to let them achieve energies up to the knee, a CR spectral feature located around $10^{15}$~eV~($=1$~PeV). Such a criticality constitutes the major theoretical hindrance for assessing SNRs as major PeV CR sources, leading to the realization that only peculiar explosions may be capable of achieving such extreme energies \citep{Bell+2013,crisotfari2020}. Specifically, these are represented by very energetic core-collapse supernovae (CCSNe), whose shocks propagate in the slow and dense medium shaped by their progenitor star's wind. The occurrence of a fast shock expanding inside a dense medium is necessary to guarantee a strong CR current at the shock that is able to trigger Bell's instability, which in turn efficiently amplifies the local magnetic field and allows particles to reach energies of $\sim$~PeV \citep{bell2004, Bell+2013}. However, the complex non-uniform environments resulting from a stellar wind expanding in the interstellar medium (ISM) strongly impact the shock evolution by producing asymmetric expansion profiles, with shocks travelling at different speeds in different portions of the ISM, depending on whether they have been expanding in cavities or impacting off interstellar clouds, making any modelling highly dependent on the specificity of individual systems. 

At the same time, indisputable observational evidence that SNRs can accelerate particles up to PeV are still missing. Recent ultra-high energy (UHE) gamma-ray (i.e. with $E>100$~TeV) observations by the Large High Altitude Air Shower Observatory (LHAASO) concerning SNRs in the northern sky have revealed two interesting middle-aged sources, IC~443 and W51~C \citep{lhaasoIC443,lhaaso_W51}. These are among the most notable examples of interacting SNRs with CCSNe progenitors. Broadband gamma-ray photons detected in these systems have been claimed to be the result of hadronic interactions, starting from GeV energies, where both remnants show evidence of the $\pi^0$-bump feature characteristic of proton-proton collisions and involving CRs with maximum energies of at least $300$~TeV and $400$~TeV, respectively. 

Ultra-high energy gamma-ray observations from these systems are thus often interpreted as evidence that SNR shocks can accelerate CRs to sub-PeV energies, as required by the standard paradigm for Galactic CR origin. This naive conclusion, however, conflicts with our general understanding of how SNRs and particle acceleration operate and therefore requires a critical assessment within the context of dedicated models, accounting for the spatial and temporal evolution of both SNR shocks and particle transport.
In fact, from a theoretical perspective, the very same process allowing PeV particle acceleration in favourable environment conditions only works efficiently within the early stages of the evolution of SNRs \citep{Bell+2013, Cardillo+2015, crisotfari2020}. In contrast, evolved systems such as middle-aged SNRs are not expected to still operate as powerful particle accelerators several thousand years after the original stellar explosion, as is testified by the steep gamma-ray spectra and the low-energy spectral breaks observed in almost all of these systems. In \citet{celli2019}, these two features are interpreted within a particle transport model specifically developed for evolved SNR shocks as being the result of a time-dependent leakage of energetic protons from the acceleration region. Specifically, the energy break corresponds to the maximum energy of confined particles, while the steep spectrum above the break is connected to the purely diffusive propagation domain of escaping particles still propagating in the remnant. In addition, the escaping of particles from evolved SNRs may also be responsible for the peculiar shape of CR electrons observed at  Earth, for example, the break observed at $E\simeq 1$\,TeV discussed in \cite{Morlino-Celli:2021}.

The process of particle escape is attracting progressively more attention in the community, as it represents the missing piece of information for a fully developed particle acceleration theory. At the very same time, this process has often been invoked to explain UHE gamma-ray observations of SNRs and their surroundings, including W51~C and some other recently discovered UHE-emitting SNRs \citep{lhaaso2026}, where multiple components in gamma-ray images and spectra were measured and putatively addressed to escaping particles illuminating gas targets. While investigating gas maps in these systems is of the utmost importance to probe the actual feasibility of hadronic scenarios, we are compelled to highlight a major shortcoming of the simplified multi-component interpretations, where the population of escaping particles is treated as a mere additional particle population component whose spectrum is left free to vary in fitting procedures independently of the confined particle population. As discussed in \citet{celli2019}, these two particle components are strongly interconnected in that the confined distribution sets the initial condition (at the respective escape time of each particle momenta) for the free-diffusion evolution of the escaping population.  

We therefore here proceed to investigate the role of IC~443 and W51~C SNRs as particle accelerators across their entire history in order to understand whether they might have ever behaved as PeVatrons at some stage of their evolution though they are clearly not PeVatrons today. We adopt a tailored particle transport model, first presented in \citet{celli2019}, that consistently embeds escaping particles in the dynamics of middle-aged SNRs. The model is further improved in this work by treating the full hydrodynamical evolution of non-radiative SNRs rather than the pure Sedov-Taylor (ST) solution as done before. Specifically, the shock evolution follows the description of \citet{truelove1999} for the time in between the two self-similar solutions of the ejecta-dominated and the deep ST phase. The particle transport solution is calculated in a uniform density profile of the circumstellar medium (CSM).

This paper is organized as follows. In Secs.~\ref{sec:ic443} and \ref{sec:w51c} we summarize the features of the two middle-aged SNRs studied here, while the details of the model adopted for the particle transport are summarized in Sect.~\ref{sec:model}. The resulting hadronic radiation is computed and presented in Sect.~\ref{sec:results}, where comparison with available gamma-ray observations is further performed to constrain the model. We present our conclusions in Sect.~\ref{sec:conclusion} about the role of IC~443 and W51~C middle-aged remnants as PeV CR accelerators.

\section{IC~443} 
\label{sec:ic443}
IC~443 (also known as the Jellyfish Nebula or G189.1+3.0) is a mixed morphology SNR \citep{rho1998} featuring a two-shell structure in the optical and radio wavelengths \citep{leahy2004} plus a thermal and centrally peaked X-ray emission, mostly arising from swept-up ISM material, as well as an X-ray bright shell \citep{troja2006} with a ring-shaped emission in hard X-rays enclosing the pulsar CXOU J061705.3+222127 \citep{olbert2001} and its pulsar wind nebula (PWN; \citealt{troja2008}). Such a complex structure is thought to have resulted from the expansion of the remnant with inhomogeneous surroundings. The remnant is indeed interacting with a dense molecular cloud \citep{burton1988} in its southern rim, producing a decelerated shock speed there, as also testified by the detection of OH maser emission \citep{hewitt2006} and X-ray spectral lines \cite{Teresi+2026cosp}. The remnant is furthermore interacting with a less dense HI cloud in its northern rim. 
Spectroscopic parallax measurements derived a remnant distance of 1.5~kpc \citep{fesen1980}, which was confirmed by \citet{welsh2003}. We adopt this value as a reference in the rest of this work. 

The estimated age of IC~443 remains a significant subject of debate, ranging from approximately 3000 to 30000 yr, depending on the specific observation and modelling techniques used. Early X-ray analyses based on shock temperatures and Sedov dynamics suggested a young age of 2800 to 3400 yr \citep{petre1988}. More recent X-ray studies of metal-rich ejecta rings proposed an age of roughly 4000 yr and argued that the remnant is expanding within a low-density cavity \citep{troja2008}. On the other hand, kinematic ages based on the proper motion of the associated neutron star suggest an older age, between 20000 and 30000 yr \citep{lee2008}. However, advanced 3D hydrodynamic modelling incorporating the interaction between the blast wave and the complex surrounding clouds has provided a different estimation of approximately 8000 years \citep{ustamujic2021}.
Significant uncertainties in these estimations arise from the different assumptions inherent in each detection method. Dynamical and plasma ages rely on X-ray measurements of temperature and ionization timescales, which are sensitive to unknown factors such as the volume filling factor and precise ambient density profiles \citep{suzuki2021}. For example, age may be underestimated if the adopted X-ray temperature were biased towards the hot interior rather than the current shock velocity \citep{lee2008}. The kinematic method, in turn, requires identifying the exact explosion centre, which can be offset from the remnant's geometric centre due to, for example, an asymmetric explosion \citep{suzuki2021}. 

In the following, we adopt the remnant age derived via the most recent 3D hydrodynamic model of \citet{ustamujic2021}, which successfully reproduced the observed X-ray morphology by incorporating the interaction between the blast wave and a toroidal cloud while setting the explosion centre at the observed PWN location. In this context, soft X-rays would arise from the interaction with surrounding clouds, while hard X-rays are dominated by shocked ejecta heated by the reflected shock. The best-fit model characterizes the parent supernova explosion energy as $E_0=1 \times 10^{51}$~erg and an ejecta mass of $M_{\rm ej} \sim 7 M_\odot$. 

The strong interaction with surrounding dense molecular clouds characterising this middle-aged remnant provides us with a key case study for hadronic gamma-ray emission. Indeed, high-energy observations by the satellite-based experiments AGILE \citep{agileIC443} and \textit{Fermi}-LAT \citep{fermiIC443} have provided independent landmark evidence of the pion-bump feature at energies below ~200 MeV \citep{fermiPB2013}, a unique spectral feature signature of neutral pion decay (produced when high-energy protons collide with ambient gas). These observations therefore provided the first direct evidence that SNRs accelerate CR hadrons. Ground-based Cherenkov telescopes have further characterized the very high energy (VHE) emission. Both the MAGIC and VERITAS observatories reported TeV emission spatially coincident with the densest molecular clouds and the site of OH maser emission \citep{magicIC443,veritasIC443_2009}. Deep observations by VERITAS over more than 150 hours have shown evidence of extended emission tracing the SNR shell and interaction regions, though the spectrum in this range is notably soft, with a slope near $\Gamma \sim -3$ \citep{veritasIC443_2015ICRC}. Despite being located in the northern sky, H.E.S.S. recently reported successful observations of the remnant with 11.4 hours of data-taking \citep{mitchell_ic443}. Detection by the HAWC observatory confirmed the presence of a point source associated with IC~443 up to 30~TeV \citep{hawc_ic4432025}, with an additional extended component (HAWC~J0615+2213) putatively related to the TeV halo of the PWN of the compact object CXOU~J061705.3+222127. Finally, the most recent and highest energy observations from LHAASO have resolved the emission into two distinct components \citep{lhaasoIC443}: i) a point-like source (dubbed C0) morphologically consistent with the $\pi^0$ source detected by \textit{Fermi}-LAT and MAGIC and characterized by a spectrum extending beyond 30 TeV without an apparent cut-off and ii) an extended source (dubbed C1) whose centroid and extension are different from HAWC~J0615+2213 while spatially overlapping with the nearby SNR G189.6+3.3. 

A simultaneous fit to the spectrum and morphology of LHAASO C0's source performed under the hypothesis of hadronic origin of the observed gamma-ray radiation has set a lower limit to the maximum energy of accelerated protons of $\sim 300$~TeV at a 95\% confidence level. \citep{lhaasoIC443}. Because the inferred value of particle maximum energy does not account for the history of the remnant, the quoted energy value should simply be regarded as the highest observed energy of particles currently interacting in the remnant. Inference about the absolute maximum energy of particles produced by the remnant blast wave at this very moment or in the past can only be obtained in the context of a detailed time-dependent model of particle transport in the remnant itself, such as the one we developed in \citet{celli2019}. We therefore proceed to apply this model while aiming at shedding light on whether IC~443 is an extinct PeVatron, that is whether it could have behaved as a PeV particle accelerator in the past. 

\section{W51~C} 
\label{sec:w51c}
W51~C (also known as G~$49.1-0.1$) is also a remnant of a CCSN explosion, and it is well known for its interactions with the dense surrounding CSM \citep{park2013}, as testified by the localization of SiO emission from this region \citep{dumas2014}, presumably produced by dust grains engulfed by the primary SNR shock. The remnant belongs to the W51 giant molecular cloud (\citealt{carpenter1998}), one of the most massive complexes of our Galaxy and which also hosts several star-forming regions and embedded star clusters therein \citep{kumar2004}. Among HII regions in the W51 complex, W51~A and W51~B show episodes of star formation at different stages, the former revealing young protostars, while the latter hosts more evolved clusters \citep{ginsburg2015}. The possibility that collective winds blown by the cluster member stars could accelerate particles up to PeV energies has been recently suggested by several authors \citep{morlino2021,vieu2023}, potentially producing UHE gamma-ray emission as a result of hadronic collisions of the accelerated particles. Specifically, several young clusters were localized in the W51 region, and their maximum energies are consistent with PeV particle production \citep{lhaaso_W51}. Whether this constitutes an alternative or complementary channel with respect to the SNR-only scenario in explaining the UHE emission observed by LHAASO is still the subject of investigation, and it requires detailed modelling of the particle acceleration and transport in the clusters, which is currently ongoing and will be published as a separate contribution (Padilha et al. 2026, in preparation).

This present work instead focuses on the remnant itself. The age and distance of the W51~C SNR have been subject to significant revision over time due to the improvements of observational techniques and evolutionary models. Early studies typically assumed a distance of $6.0$~kpc based on molecular-line observations that placed the remnant behind a ridge of molecular gas. At this distance, using a standard Sedov model for the adiabatic expansion phase (while neglecting the effect of ejecta), one can derive an age of approximately 30~kyr \citep{koo1995,sasaki2014}. Some authors \citep{zhang2017,reyes2022} have revised the remnant distance down to values between $4.1$~kpc (based on the systemic velocity of newly discovered optical filaments) and $4.3$~kpc (via H I absorption spectroscopy), which lead to smaller physical radii and consequently younger age estimations, such as 13~kyr. More recently, the remnant distance was instead assessed to be at 5.4~kpc \citep{ranasinghe2018}. Refined hydrodynamic models following \citet{truelove1999} evolution with a $20M_\odot$ progenitor imply an age of 18~kyr \citep{leahy2018}. These models are more accurate than the early Sedov calculations in \citet{koo1995}, as they account for the transition from the free expansion phase to the adiabatic phase and recognise that the remnant's expansion is still influenced by its initial stellar mass. Recent radio surveys have identified the so-called north-east edge as part of the remnant itself, thus increasing its angular diameter to 37'. As a result, additional magneto-hydrodynamic (MHD) simulations also estimated an age of 18~kyr \citep{zhang2017}. In the following, we adopt a distance of 5.4~kpc and an age of 18~kyr, values that are consistent with both the larger morphological features of the remnant and the high-mass ejecta signatures found in X-ray spectral analyses. 

The W51 region has become a primary target for studying CR accelerators due to its bright gamma-ray emission spanning nearly six orders of magnitude in energy. Observations by \textit{Fermi}-LAT first discovered extended GeV emission, identifying the characteristic $\pi^0$-bump below 1~GeV \citep{fermi_W51}. Subsequent VHE detections by H.E.S.S. \citep{hess_w51} and MAGIC \citep{magic_w51} confirmed this emission in the TeV range, showing that the radiation is spatially coincident with the interaction region between the SNR shock and dense molecular gas. However, MAGIC data show evidence of a gamma-ray source composed of two components above 1 TeV,  which are not yet confirmed by H.E.S.S. One component is coincident with the interaction region between W51~C and W51~B, while the other is coincident with the potential PWN CXOU~J192318.5+140305 \citep{koo2005}, suggesting that HESS~J1923+141 emission may result from a mix of different astrophysical objects \citep{hgps}. Recent UHE observations by LHAASO have extended the measured spectrum up to 200 TeV \citep{lhaaso_W51}, suggesting an exponential cut-off in the parent proton spectrum at approximately 400 TeV. As for IC~443, we remark that interpreting this result as the maximum energy of the accelerator is incorrect in that a physical model is needed to interpret the observations and infer the value of such a parameter and its dynamical evolution. Models coupling the particle to the SNR evolution already exist (see \citet{celli2019}) and can be tested straightforwardly, as we do in the next sections.

\begin{table*}
\caption{Values of parameters used to model gamma-ray emission from the two SNRs in the uniform ambient medium case ($n_0=10 \, {\rm cm}^{-3}$). }
\label{table:1}     
\centering                          
\begin{tabular}{c|c c c c c | c c c c c | c }        
\hline              
Source & $E_{\rm SN}$   &   $M_{\rm ej}$   &   $t_{\rm SNR}$   &   $d$  &   $t_{\rm Sed}$ &   
$\alpha$  &    $\xi_{\rm CR}$  &   $p_{\rm M}c$  &   $\delta$   &   $p_{\rm max}(t_{\rm SNR})c$  &    $D/D_{\rm Gal}$   \\    
\hline                       
 IC 443 & $10^{51}$ erg  &  $ 7\,{\rm M}_{\odot}$ & 8 kyr & 1.5 kpc & 551 yr & 4.0  &  1\% &  $\sim 100$\,TeV  &  3  & $\sim 33$\,GeV & 0.3  \\
 ''     &  ''            &           ''           &   ''  &   ''  &   ''  & 4.3  &  2\% &  $\sim 300$\,TeV  &  3  & $\sim 98$\,GeV & 0.3  \\
 \hline
 W51 C  & $10^{51}$ erg  &  $ 10\,{\rm M}_{\odot}$ & 18 kyr & 5.4 kpc & 741 yr &  4.0 & 10\% & $\sim 200$\,TeV  &  3  & $\sim 14$\,GeV & 0.1 \\
   ''   &   ''           &       ''               &    ''  &   ''  &   ''  &  4.3 & 12\% &  $\sim 500$\,TeV  &  3  & $\sim 35$\,GeV & 0.1 \\
\hline                                   
\end{tabular}
\tablefoot{The first block refers to the SNR properties, the second to the acceleration and escaping properties, and the last one to the diffusion coefficient of the surrounding medium normalized to the average Galactic one. Note that $p_{\max}(t_{\rm SNR})c$ refers to the maximum energy at the current SNR age.}
\end{table*}

\section{Modelling particle acceleration and escape}
\label{sec:model}
The theoretical description of particle acceleration and propagation usually adopted for SNR spectral modelling generally suffers from severe limitations, the major ones being adopting one-zone models for particle interactions and neglecting the dynamics of shocks. Both assumptions affect predictions of the shape of the energy spectra of secondary gamma rays and neutrinos. In \citet{celli2019}, we developed a phenomenological model for particle production in SNR shocks, including the temporal evolution of the forward shock in adiabatic expansion as well as the dynamics of the particle acceleration and injection into the ISM. Analytical solutions for the particle distribution function were derived under the assumptions of a uniform ISM, spherical symmetry, and CR stationary acceleration efficiency ($\xi_{\rm CR}$). Analytical solutions to the problem were found for power-law (in momentum) acceleration spectra of particles of different specific slopes, namely $f_0(p) \propto p^{-4}$ and $f_0(p) \propto p^{-4.3}$, the former being the standard prediction for test-particle DSA theory, while the latter represents steeper injection spectra possibly produced as a result of non-linear CR feedback \cite[see, e.g.,][]{Caprioli+2020}. 
We notice that a similar approach has been adopted in other works, such as \cite{Brose+2020, Brose+2021}, with a fully numerical approach for the solution of the transport equation and SNR expansion.

In \citet{celli2019}, the maximum particle momentum, $p_{\rm max}(t)$, is assumed to increase linearly during the ejecta-dominated phase and then decrease as a power-law during the ST phase:
\begin{equation} 
\label{eq:pmax}
 p_{\max}(t) =
  \begin{cases} 
   p_\textrm{M} \left( t/t_{\rm Sed} \right)     & \text{if } t \leqslant  t_{\rm Sed}    \\
   p_\textrm{M} \left( t/t_{\rm Sed} \right)^{-\delta}     & \text{if } t > t_{\rm Sed} \,,
  \end{cases}
\end{equation}
where $\delta$ is a free parameter of the model that encodes the temporal evolution of the upstream magnetic turbulence. Inverting this relation, the escape time of particles with momentum $p$ is defined as 
$$t_{\rm esc}(p)=t_{\rm Sed}\left( \frac{p}{p_{\rm M}} \right)^{-1/\delta} \, , $$
where $p_{\rm M}$ represents the absolute maximum momentum achieved at the Sedov time ($t_{\rm Sed}$), which can be parametrised as
\begin{equation}
   t_\textrm{Sed} \simeq 1.6 \times 10^3 \, \textrm{yr} \left(\frac{E_\textrm{SN}}{10^{51} \, \textrm{erg}} \right)^{-1/2} 
                \left(\frac{M_\textrm{ej}}{10 \, M_\odot} \right)^{5/6} 
                \left(\frac{\rho_0}{1 \, m_\textrm{p}/\textrm{cm}^3} \right)^{-1/3} \, ,
\end{equation}
depending on the upstream density $\rho_0$, the explosion energy $E_{\rm SN}$, and the ejecta mass $M_{\rm ej}$. Here, $m_{\rm p}$ is the proton mass. The adopted maximum energy prescription implies that particle escape is energy-dependent, with higher-energy particles leaving the shock earlier, at a rate regulated by $\delta$. Such a parameter depends primarily on the mechanism producing the magnetic field turbulence and can be estimated in the time-limited acceleration scenario, as described in Appendix~A of \citet{celli2019}. During the Sedov-Taylor stage, $\delta$ is expected to approach the value of $1/5$ if no magnetic field amplification occurs, the value of $7/5$ if the field amplification is due to the resonant streaming instability only, and the value of $2$ if the non-resonant streaming instability dominates the amplification. Additional effects, such as magnetic damping or MHD-related processes, might produce larger values than these simple analytical estimates.

As such, at each time $t$, the particle population inside the SNR is described as consisting of two components:  
\begin{itemize}
\item[(i)] Confined particles (located only inside the SNR), with $t_{\rm esc}(p) > t$, which are tightly coupled to the expanding plasma and undergo adiabatic losses. 
\item[(ii)] Non-confined (or escaping) particles (located both inside and outside the SNR), with $t_{\rm esc}(p) < t$, which have decoupled from the shock turbulence and evolve under pure spatial diffusion described by a spatially homogeneous diffusion coefficient $D(p)= \chi D_{\rm Gal}(p)$, where $\chi \leq 1$ quantifies the suppression of diffusion relative to the average Galactic value for which we consider a Kolmogorov-like parametrisation as $D_{\rm Gal}(p)=10^{28}(p/10 \, {\rm GeV/c})^{1/3}$~cm$^2$/s.
\end{itemize}

The determination of the total particle distribution function, provided by the sum of confined and escaping hadrons, is affected by the temporal evolution of the SNR forward shock. Similar to our previous work \citet{celli2019}, we here investigate the case of a spatially uniform ambient medium, fixing the upstream number density to typical giant molecular cloud values, i.e. $n_0=10$~cm$^{-3}$.

\section{Hadronic non-thermal radiation}
\label{sec:results}

The computation of hadronic collisions requires defining the radial profile for the shocked target gas inside the SNR, where most of the secondaries are expected to be produced because of the density compression effect resulting from the shock passage. In the uniform CSM case, we considered the following polynomial parametrisation for $r<R_{\rm sh}$, which well approximates the Sedov solution:
\begin{equation}
\label{eq:sedovDensity}
  n_{\rm in}(t,r) = n_0 \sigma \left[ a_1 X^{\alpha_1} + a_2 X^{\alpha_2}  + a_3 X^{\alpha_3} \right] \, .
\end{equation}
Here, $X= r/R_{\rm sh}(t)$, $n_0=\rho_0/m_{\rm p}$, $\sigma$ is the compression ratio at the shock (which is connected to the accelerated particle slope), and the parameters $a_1, a_2,a_3$ and $\alpha_1, \alpha_2, \alpha_3$ are provided in \citet{celli2019} as a fit of the original Sedov solution. The computation of gamma rays from hadronic collisions is here performed by convolving the differential energy spectrum of protons residing in the remnant interior with the density profile of Eq.~\eqref{eq:sedovDensity} and considering the differential cross-section for pp collisions, parametrised following \cite{kafexiu2014} with SIBYLL 2.1 values. 

Concerning the particle escape model, we here exploit a limited number of values for the unknown model parameters, selected based on previous indications from \citet{celli2019}. Specifically, we test values of $\delta$ in the range $[1,4]$ in steps of one and diffusion coefficient normalisation values with discrete values of $\chi$ in $[0.01, 0.03, 0.1, 0.3]$. We note that larger values of $\delta$ imply a faster particle escape from the shock region and result in a lower energy break in the hadronic gamma-ray emission produced by the particles interacting within the remnant itself and a more intense emission from its outer regions, as shown in Fig.~5 of \citet{celli2019}. We derive the CR acceleration efficiency by fitting the absolute normalisation of the multi-band gamma-ray data from each source via a $\chi^2$ minimization procedure while also accounting for upper limits set by different experiments that are treated at a 95\% confidence level in a one-sided Gaussian convention.

\section{Results}

\begin{figure*}
\centering
\includegraphics[width=0.48\textwidth]{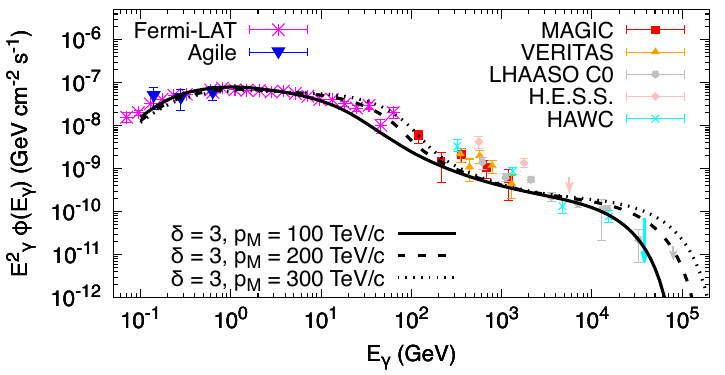}
\includegraphics[width=0.48\textwidth]{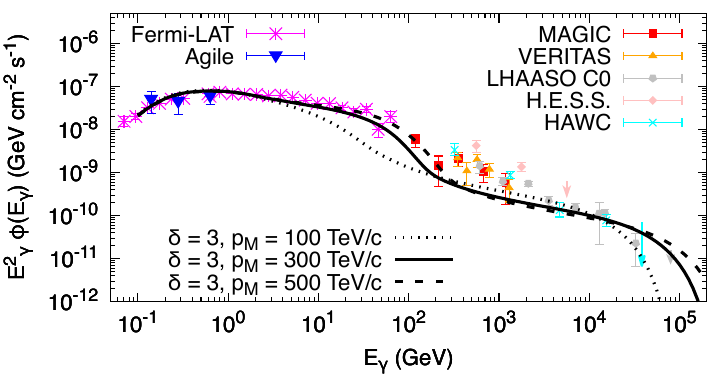}
\caption{IC~443 modelling in uniform CSM. \emph{Left}: $f_0 \propto p^{-4}$ acceleration spectrum at forward shock. \emph{Right}: $f_0 \propto p^{-4.3}$. Best-fit values obtained for each model are shown as solid lines and given in Table~\ref{table:1}. A factor of three suppression of the diffusion coefficient compared to standard ISM values is required in both cases. Data from: \textit{Fermi}-LAT \citep{fermiIC443}, Agile \citet{agileIC443}, MAGIC \citet{magicIC443}, VERITAS \citet{veritasIC443_2009}, H.E.S.S. \citet{mitchell_ic443}, HAWC \citet{hawc_ic4432025}, and LHAASO \citet{lhaasoIC443}.}
\label{fig:ic443_uniform}
\end{figure*}

\begin{figure*}
\centering
\includegraphics[width=0.48\textwidth]{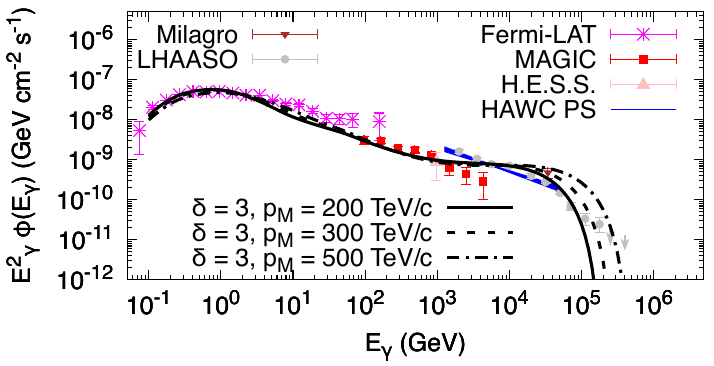}
\includegraphics[width=0.48\textwidth]{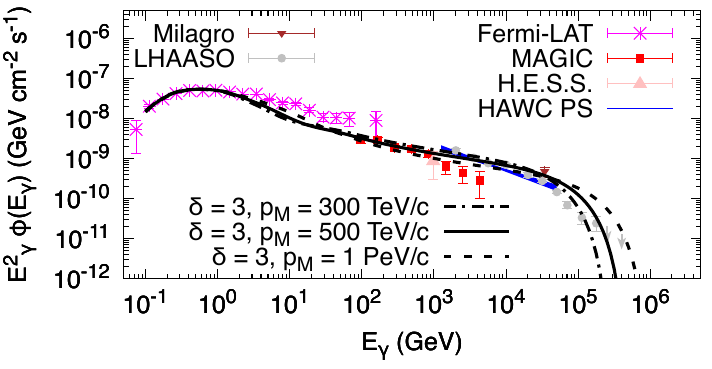}
\caption{W51~C modelling in uniform CSM. \emph{Left}: $f_0 \propto p^{-4}$ acceleration spectrum at forward shock. \emph{Right}: $f_0 \propto p^{-4.3}$. Best-fit values obtained for the models are shown as solid lines and given in Table~\ref{table:1}. A factor of ten suppression of the diffusion coefficient compared to standard ISN values is required in both cases. Data from: \textit{Fermi}-LAT \citet{fermi_W51}, MAGIC \citet{magic_w51}, H.E.S.S. \citet{hess_w51}, HAWC \citet{hawc_w51}, LHAASO \citet{lhaaso_W51}, Milagro \citet{milagro_w51}.}
\label{fig:W51_uniform}
\end{figure*}

For both SNRs, we consider the same value of kinetic energy, $E_{\rm SN}=10^{51}$~erg, which is consistent with the standard values of kinetic energy release in CCSNe. We fix the age according to literature. We further assume uniform ejecta profiles, whose total masses are reported in Table~\ref{table:1} according to results from MHD simulations and literature studies of the remnant progenitors, tuned to reproduce the observed remnant size, as presented in Secs.~\ref{sec:ic443} and \ref{sec:w51c}.

Figure~\ref{fig:ic443_uniform} provides the best-fit models to the gamma-ray spectrum of IC~443 when assuming evolution in a uniform CSM. Among all the tested models, the best-fit solution requires a mild suppression of the local (both internal and nearby) diffusion coefficient by a factor of three compared to the average ISM value. For such a value, we show in the two panels models obtained with either a $f_0(p) \propto p^{-4}$ pure acceleration spectrum (left panel) or a steeper acceleration spectrum as $f_0(p) \propto p^{-4.3}$. It is interesting to note how our model naturally reproduces the spectral curvature of data around 100~GeV photon energies, thanks to the features of the escape process, namely the introduction of a natural energy break and spectral steepening above it. The best-fit values for the  standard DSA scenario indicate that the combination of $\delta=3$ and $p_{\rm M} \lesssim 200$~TeV/c could reasonably well reproduce gamma-ray data under the assumption of fully hadronic origin of the radiation without violating any available flux upper limit. The amount of interactions between the particles currently inside the remnant shock (both confined and escaping) tends to indicate a CR acceleration efficiency of 1\% across the remnant's history in case of a uniform CSM of $n_0=10$~protons/cm$^3$. Compared to the case of a  steeper acceleration spectrum as $f_0(p) \propto p^{-4.3}$, we find a stable value of $\delta$, with a slightly higher CR acceleration efficiency (well within the standard 10\% assumption), while a larger  maximum energy of $p_{\rm M}\lesssim 300$~TeV/c (right panel) could be allowed. The best-fit values of the two solutions are given in Table~\ref{table:1}.

Similarly, Fig.~\ref{fig:W51_uniform} provides the best-fit models to the gamma-ray spectrum of W51~C, assuming evolution in the uniform CSM scenario. In this case a higher suppression of the diffusion coefficient is required to explain its gamma-ray spectrum compared to IC~443. In fact, the best-fit solutions of both the $f_0(p) \propto p^{-4.0}$ and $f_0(p) \propto p^{-4.3}$ models are found with a factor of ten smaller diffusion coefficient than the average ISM. The best representative model of the broadband gamma-ray spectrum in the linear DSA acceleration scenario (shown in the left panel) requires $\delta=3$ jointly with a maximum particle momentum at Sedov time of $p_{\rm M}\lesssim 300$~TeV/c and a CR acceleration efficiency of 10\%. With the steeper acceleration spectrum, a slightly larger efficiency of up to 12\% and $p_{\rm M}\lesssim 500$~TeV/c are in turn allowed. 
Even though the absolute maximum energies at Sedov times obtained for W51~C are higher than for IC~443, showing that the former is a more extreme accelerator than the latter (as further confirmed by the fitted CR conversion efficiencies), none of them appear to have ever achieved PeV energies, which is required to justify CRs up until the knee. 

We remark that the CR acceleration efficiencies listed in Table~\ref{table:1} were derived assuming that a constant fraction of the shock ram pressure is converted into non-thermal particles over time. The derived values are partially degenerate with the assumed external gas density. Since the product of these two quantities sets the rate of hadronic collisions, a higher gas density would require a correspondingly lower acceleration efficiency to reproduce the observed gamma-ray flux. This degeneracy is particularly relevant for W51~C, whose adopted upstream numerical density is lower than the value considered in other studies \cite[see, e.g.,][]{lhaaso_W51}. We stress that the adopted density value should be read as a reference value for the target gas rather than as a measured mean density capturing the multi-phase interstellar gas around these remnants.

The same density, however, also sets the expansion rate of the forward shock, and a single value cannot simultaneously satisfy both roles. Adopting $n_0=10$~cm$^{-3}$ for the shock dynamics and the remnant ages of Table~\ref{table:1}, the \citet{truelove1999} solution predicts forward-shock radii of $\sim 7.7$~pc for IC~443 and $\sim 11$~pc for W51~C, which is appreciably smaller than the observed radio sizes of $\sim 10$~pc and $\sim 24$~pc, respectively \citep{Green:2025}.\footnote{https://www.mrao.cam.ac.uk/surveys/snrs/snrs.data.html} Matching the observed sizes instead requires substantially lower densities of 2.5~cm$^{-3}$ and 0.15~cm$^{-3}$ for IC~443 and W51~C, respectively. 
This inconsistency is probably due to a radially dependent density profile whose dynamical densities, i.e. the values effectively swept up by the forward shock on average over the remnant's lifetime, do not coincide with the target density relevant for hadronic gamma-ray production, with the latter instead tracing the denser gas encountered at later times when the shock impacts a molecular cloud. 

Such a distinction is physically motivated, as both remnants originate from CCSNe, whose massive progenitors are expected to carve out a lower-density cavity through their pre-supernova winds. This is directly supported in the case of IC443 by the low-density cavity inferred from the X-ray study of \citet{troja2006}, while the interaction with much denser gas is independently evidenced by the OH maser emission detected towards IC~443 \citep{hewitt2006} and the SiO emission detected towards W51~C \citep{dumas2014}. Introducing a wind-like radial density profile in the particle-escape framework would capture this dichotomy self-consistently, but this requires a substantial extension of the analytical treatment of \citet{celli2019} or, alternatively, the use of a full numerical description, which is left to future work.

Finally, we recall that our model assumes a constant acceleration efficiency through the parameter $\xi_{\rm CR}$. Such an assumption is motivated by the roughly constant acceleration efficiency inferred from particle-in-cell (PIC) simulations when the shock Mach number is $M \gtrsim 5$ \citep{Caprioli-Spitkovsky:2014}. However, a temporal dependence of this parameter is expected as the shock slows down, namely after entering the Sedov stage. On a general ground, an acceleration efficiency decreasing with time, i.e $\xi_{\rm CR} \propto t^{-\eta}$ and $\eta>0$, would produce harder spectra than those obtained in our calculation for $p> p_{\max}(t)$, while the slope for $p<p_{\max}(t)$ would remain unchanged and equal to the  spectrum at the shock. Given that in our present model the steepening of the spectrum for $p> p_{\max}(t)$ 
is determined by the time behaviour of the maximum energy (hence by the parameter $\delta$), the introduction of a time dependent efficiency would result in a degeneracy in the parameters ($\eta$, $\delta$). 
However, because the very same PIC simulations \citep{Caprioli-Spitkovsky:2014} are compatible with $\xi_{\rm CR} \sim M^{0.2} \sim t^{-0.1}$, the model predictions for $\delta$ would only be marginally affected.

\section{Conclusions}
\label{sec:conclusion}
We have investigated whether IC~443 and W51~C could have behaved as PeV CR accelerators at some point in their evolution through a self-consistent treatment of particle escape within the dynamics of the forward shock. We applied this model to the observed full broadband gamma-ray spectra of the two remnants, including the recently reported UHE components. In both cases, we found a purely hadronic origin of the radiation reproduces the data well, with the escaping particle population naturally accounting for the spectral steepening observed above $\sim 100$~GeV without invoking ad hoc multi-component fits.

In both cases, the best-fit solutions favour a particle maximum-energy that decreases in time as $t^{-3}$, regardless of the assumed acceleration slope. This slope is steeper than the ones typically adopted in simplified escape prescriptions, suggesting that a faster decline of the maximum energy is required compared to the one predicted by the pure Bell-type field amplification. At the same time, the required diffusion suppression (a factor of three to ten relative to the average Galactic value) and the inferred CR acceleration efficiencies (1–10\% in linear DSA theory, up to 12\% for the case where non-linear effects are included) both fall within physically reasonable ranges, supporting the internal consistency of the model and the overall SNR paradigm for particle acceleration.
We highlight that the suppressed diffusion immediately outside the SNRs seems to be a quite common feature, as similar values have been inferred for W44 \citep{W44-MAGIC:2025} and $\gamma$-Cygni SNR \citep{g-Cygni_MAGIC:2023}. Such a suppression requires an enhanced level of magnetic turbulence, which may be either due to local pre-existing turbulence or the result of the very same escaping particles that trigger the resonant \citep{Nava+2016, DAngelo+2018} and non-resonant streaming instability \citep{Schroer+2022} outside the SNR.

Nonetheless, across all parameter configurations exploited here, the maximum particle momentum at the Sedov time remains in the range of 100–500 TeV/c and hence systematically below ~1 PeV/c. We conclude that IC~443 and W51~C, though consistent with efficient particle acceleration and confinement, show no signatures of PeVatron activity, neither active nor extinct. 

As a final remark we warn the reader that the regions we analysed are quite complex, and the contribution from other sources to the gamma-ray emission cannot be completely excluded, especially for the W51 region, where several young star clusters are present and may be responsible for the highest energy emission (Padilha et al. 2026, in prep.). As a consequence a more detailed analysis accounting also for the spatial morphology is highly desirable. The forthcoming CTAO \citep{CTAO-book:2019} and ASTRI \citep{ASTRI-core_science:2022, ASTRI-Gal_science:2022} facilities, thanks to the their superior angular resolution, will improve the accuracy of this type of study in the very near future.

\begin{acknowledgements}
SC gratefully acknowledges support from the “Award Horizon Europe 2025” funding scheme by Sapienza Università di Roma under grant ID AH1251992EC2A31C.
GM is partially supported by the INAF Theory Grant 2024 {\it Star Clusters As Cosmic Ray Factories II}.
\end{acknowledgements}

\bibliographystyle{aa} 
\bibliography{references} 

\end{document}